\documentclass{article}

\usepackage[preprint]{spconf}
\copyrightnotice{978-1-6654-5227-4/22/\$31.00~\copyright2022 IEEE \hfill}
\usepackage{amsmath,graphicx}

\usepackage{url}

\usepackage{amssymb,amsmath,epsfig}
\usepackage{cite}
\usepackage[bookmarks=false]{hyperref}
\usepackage{graphicx}
\usepackage{booktabs}
\usepackage{multirow,multicol}
\usepackage{subcaption}
\usepackage{balance}
\usepackage{amssymb,epsfig}
\usepackage{booktabs}
\usepackage{multirow,multicol}
\usepackage{subcaption}
\usepackage{color}
\usepackage{multirow}

\usepackage[utf8]{inputenc}
\usepackage[T1]{fontenc}

\usepackage{color}

\usepackage{wrapfig}
\usepackage{xurl}

\usepackage{mathtools}

\title{Frequency-centroid features for word recognition of non-native English speakers}

\name{Pierre~Berjon$^{\star1}$, Rajib~Sharma$^{\star2}$,~Avishek~Nag$^{3}$,~Soumyabrata~Dev$^{4,5}$
\thanks{$^{\star}$ Joint first authors; both authors contributed equally.}
\thanks{This research was conducted with the financial support of Science Foundation Ireland under Grant Agreement No.\ 13/RC/2106\_P2 at the ADAPT SFI Research Centre at University College Dublin. ADAPT, the SFI Research Centre for AI-Driven Digital Content Technology, is funded by Science Foundation Ireland through the SFI Research Centres Programme.}
\thanks{Send correspondence to S.\ Dev, email: soumyabrata.dev@ucd.ie}
}
\address{$^{1}$ Department de Sciences du Numérique, INP-ENSEEIHT, Toulouse, France \\
         $^{2}$ Department of DSIS, Indian Institute of Information Technology Dharwad, India \\
         $^{3}$ School of Electrical and Electronic Engineering, University College Dublin, Ireland \\
         $^{4}$ ADAPT SFI Research Centre, Dublin, Ireland \\
         $^{5}$ School of Computer Science, University College Dublin, Ireland}

\begin{document}

\maketitle

\begin{abstract}
The objective of this work is to investigate complementary features which can aid the quintessential Mel frequency cepstral coefficients (MFCCs) in the task of closed, limited set word recognition for non-native English speakers of different mother-tongues. Unlike the MFCCs, which are derived from the spectral energy of the speech signal, the proposed frequency-centroids (FCs) encapsulate the spectral centres of the different bands of the speech spectrum, with the bands defined by the Mel filterbank. These features, in combination with the MFCCs, are observed to provide relative performance improvement in English word recognition, particularly under varied noisy conditions. A two-stage Convolution Neural Network (CNN) is used to model the features of the English words uttered with Arabic, French and Spanish accents.
\end{abstract}
\section{Introduction}
\label{Sec1}

Automatic speech recognition (ASR) is the technology by which sounds captured by a microphone are transcribed into written words by machines \cite{rabiner1978digital,rabiner2007introduction,benesty2008springer}. Since its inception in the 1950s, ASR has made considerable progress with the help of phoneticians, linguists, mathematicians and engineers, who defined the acoustic and linguistic knowledge necessary to fully understand human speech. The proliferation of voice-technology in the last decade has seen mobile devices such as Google Home, Amazon Alexa, Apple Siri, and other less known ones, become an indispensable part of the modern home. As such, research on keyword or ``wake-word'' recognition, which is a subset of speech recognition, has come to prominence\cite{warden2018speech,arik2017convolutional,tang2018deep,wang2019adversarial,sharma2020adaptation,zeng2019effective,lopez2019keyword,lopez2020improved}.

In spite of such progress, ASR systems struggle under many real-life challenging conditions \cite{o2008automatic,alharbi2021automatic}. The favorable conditions for ASR involve native speech belonging to a single speaker with proper diction (not presenting a voice pathology), recorded in a quiet and noiseless environment, and based on a common vocabulary (words known to the system~\cite{hossari2019test}). System performance decreases when dealing with accents of non-native speakers, different dialects, speakers with voice pathology, words unknown by the system (usually proper names), noisy audio signals (low signal-to-noise ratio), \textit{etc.} \cite{kanabur2019extensive,Reitmaier2022,Latif2020,Huang2004} Among the many challenges facing ASR, one key challenge is the accent of the speaker, especially when s/he is not fluent in the language in question. In ~\cite{berjon2021}, Berjon \textit{et al.} analyzed speech spectrograms, and identified the limitations of French idiosyncrasies on English word recognition. Indeed, the performance of ASR systems are generally inferior for non-native speakers than that of for native speakers\cite{o2008automatic,alharbi2021automatic,huang2004accent}. Naturally, just like ASR, automatic word recognition (AWR), being a subdomain of ASR, is also challenged by the accent of the speakers, particularly when the speakers do not naturally speak the language.

With the widespread application of artificial neural networks (ANNs) in recent times in different tasks~\cite{hossari2018adnet,John2021,Dev2019}, it has been successfully applied in word recognition as well \cite{warden2018speech,arik2017convolutional,tang2018deep,wang2019adversarial,sharma2020adaptation,zeng2019effective,lopez2019keyword,lopez2020improved}. The neural-networks based ASR and AWR systems have broken through the limitations of many traditional hidden Markov model (HMM) based systems \cite{Wang2018,Shrestha2019,purwins2019deep}. This does not mean, however, that we can neglect the process of feature extraction/engineering. Undoubtedly, the Mel frequency cepstral coefficients (MFCCs) remain the most important features for all speech/voice applications. However, they may not always be the optimal features, and other complementary features may be needed to boost performance in real-life conditions \cite{EMD_RS_Review,holambe2012advances,lopez2020improved,wang2018multiobjective,elshamy2018dnn,tachioka2018dnn,jung2019self}. In the case of ASR (hence AWR), it has been suggested that the reduction in performance improvement from accent-mismatch is due to the different characteristics of the language in question, and not because of the speaker itself \cite{article1,article2}. As such, the objective of this work is to explore frequency-based features from the Fourier spectra which can complement the MFCCs in the task of AWR. In this study, we consider five English words uttered by native speakers of two European (Spanish and French) and one non-European (Arabic) language. These `native' speakers consider either Spanish, French or Arabic as their first language, primary language or mother-tongue, but are multi-lingual and can also converse in English. In other words, they are `non-native' English speakers with either Spanish, French or Arabic accents. The detection of accents is an important task, as it can be readily applied to the generation of effective sub-titles in video tutorials~\cite{batra2022dmcnet} during online education. For each accent-category, a two-stage Convolution Neural Network (CNN) is utilized to classify the individual words, using a combination of our proposed features with the conventional MFCCs as the input.

The rest of this paper~\footnote{In the spirit of reproducible research, the codes related to this article is shared via \url{https://github.com/pberjon/Frequency-centroid-features-for-word-recognition-of-non-native-English-speakers}.} is structured as follows. In Section~\ref{sec:features} the proposed novel features and the CNN architecture used to implement the AWR system is discussed. Section~\ref{sec:results} presents the results, and Section~\ref{sec:conclusion} summarizes this work.

\vspace{-0.0cm}
\section{Features and Model for AWR}
\label{sec:features}
\vspace{-0.0cm}

The MFCCs are the most widely used features in speech processing. The MFCCs are derived using the Mel filterbank, which mimic the spectral analysis process of the basilar membrane of the inner ear\cite{rabiner1978digital,rabiner2007introduction,benesty2008springer}. The Mel frequency scale represents the perceptual frequency scale of the ear, wherein certain ``critical bands'' significantly influence our perception of sound\cite{rabiner1978digital,rabiner2007introduction,benesty2008springer}.

\begin{figure}[htb]
\centering
\vspace{-0.2cm}
\includegraphics[width=0.49\textwidth]{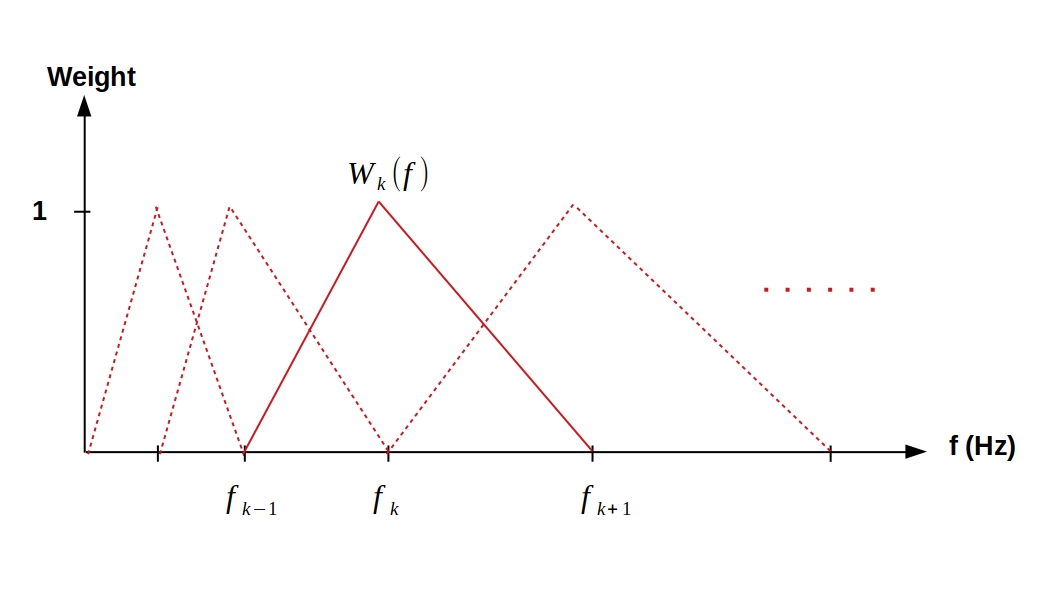}
\vspace{-0.5cm}
\caption{Block diagram of the Mel filterbank.}
\label{f_1}
\end{figure}

\begin{figure*}[t]
\centering
\includegraphics[height=3.7cm]{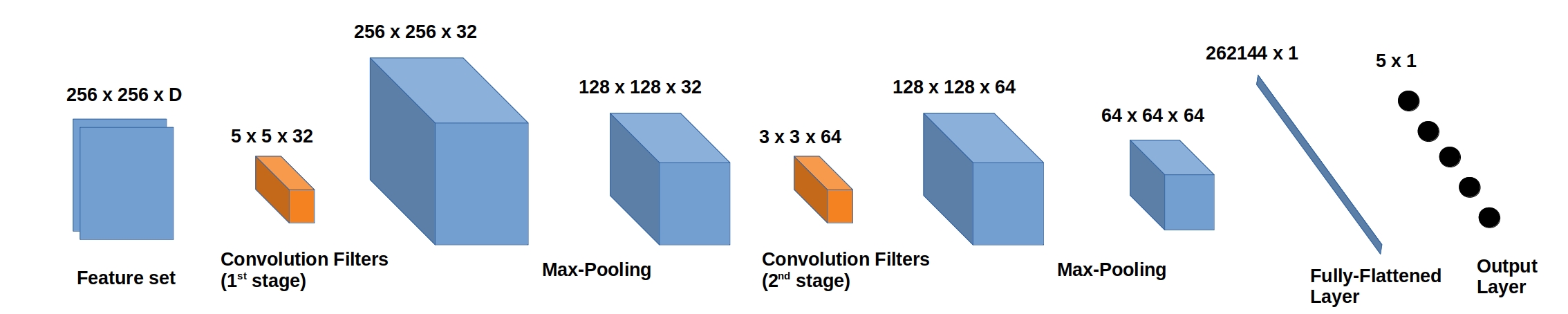}
\caption{CNN with two stages used for modelling the five words. Relu activation function is used.}
\label{f_CNN}
\end{figure*}

The Mel scale is given by
\begin{equation}
    f_{\mathrm{mel}} = 2595~\log_{10} (1+\frac{f}{700}).
\end{equation}
where $f_{\mathrm{mel}}$ represents the frequency in the perceptual frequency (Mel) scale, and $f$ represents the analog frequency (Hz). The representation of the critical bands in the analog frequency (Hz) scale results in a sequence of non-uniform and overlapping bands of triangular filters, well known as the Mel filterbank, as shown in Figure~\ref{f_1}.

Speech being a highly non-stationary signal, it is indeed processed in short frames or segments of a few tens of milliseconds. Applying the Mel filterbank on the power spectrum of the pre-emphasized speech-segment, $s'(n)$, we obtain the MFCCs in the following few steps\cite{rabiner1978digital,rabiner2007introduction,benesty2008springer}:
\begin{align}
s'(n) & = s(n) - 0.98 s(n-1) \longleftrightarrow S'(f),
\\
\hat{S'_k} &= \int_f |S'(f)|^2 W_k (f) ~df,~ k\in [1,K],
\\
\hat{C_{n}} &= \sum_{k=1}^{K} (\log(\hat{S'_k})) \cos [n(k - \frac{1}{2}) \frac{\pi}{K}],~n\in [1,K]
\end{align}
In the above equations, $K$ is the total number of filters in the Mel filterbank, $W_k (f)$ being the $k^{\mathrm{th}}$ filter. $\hat{S'_k}$ is the output after passing the power-spectrum of the pre-emphasized speech-segment, $|S'(f)|^2$, through the filter. $\hat{C_n}$ is the $n^{\mathrm{th}}$ cepstral coefficient: the compressed feature obtained by Discrete Cosine Transform (DCT).\\

As is evident, the MFCCs are purely energy-based features. In this work, we aim to utilize the frequency information of the critical bands of the Mel filterbank. As such, we propose the frequency-centroid (FC) features, given by,
\begin{align}
s(n) &\longleftrightarrow S(f),
\\
S_k(f) &= 
\left\{\begin{matrix}
S(f),~ f_{k-1} < f < f_{k+1}
\\ 
0,~\mathrm{otherwise}
\end{matrix}\right.,
\\
\color{blue}F_k &= \frac{f |S_k(f)|}{  \int_f |S_k(f)| ~df  },~ k=1,2,...,~K.
\end{align}

To model the features discussed, we propose, after experimentation, a two-stage Convolution Neural Network (CNN), as shown in Figure~\ref{f_CNN}. This architecture is far simpler that any large vocabulary continuous-speech recognition system, which includes not only an `acoustic model' but also a `language model' \cite{rabiner1978digital,rabiner2007introduction,benesty2008springer}. Our focus is on building a `small-footprint' system sufficient for a limited-set word recognition system \cite{warden2018speech,arik2017convolutional,tang2018deep,wang2019adversarial,sharma2020adaptation,zeng2019effective,lopez2019keyword,lopez2020improved}. For this purpose, multi-stage (upto 5) CNNs were experimented with, but without much gain in performance. As such, the final proposed neural-network for this task consists of two convolution layers: one with 32 filters and a ReLu activation function, and another with 64 filters and a ReLu activation function. For any particular accent, five English words are modelled, as represented in the output layer. The input layer, or feature set, consists of a two-dimensional matrix ($D = 1$ in Figure~\ref{f_CNN}) if either the MFCCs or the FCs are used as features. If both are used, the input layer is a three-dimensional matrix with $D = 2$.

\section{Experimental setup and results}
\label{sec:results}

For our experiments, we utilize a subset of the Speech Accent Archive database\footnote{\url{https://accent.gmu.edu/index.php}}. The database consists of speakers from 177 countries belonging to 214 different mother tongues, speaking a particular English passage. From the database, we have arbitrarily selected a speech subset of only three different accents (French, Spanish and Arabic), from which we have manually isolated five specific words for the isolated word recognition task. The speakers with these accents consider their `accents' as their primary language or mother-tongue, however they may be from different regions of the world. For example, the speakers with Spanish accents are constituted from not only Spain but different regions of Latin America and other former Spanish colonies. The details of the limited dataset are presented in Table~\ref{t_dataset}. On an average, the utterances in the dataset are of 0.5~s duration. Unfortunately, equal number of utterances could not be easily collected for each of the accents for any particular word. However, for any particular accent, the five different words have the same number of utterances, and form a balanced dataset.
\begin{table}[h!]
\vspace{-0.0cm}
\renewcommand{\arraystretch}{1.1}
\centering
\caption{ Number of speech signals used for training and testing for word recognition of five English words uttered by speakers with three different accents.}
\label{t_dataset}
\resizebox{0.5\textwidth}{!}{%
\begin{tabular}{|c|c||ccccc|}
\hline
\multirow{2}{*}{Accent}  & \multirow{2}{*}{Data} & \multicolumn{5}{c|}{Words}                                                                                                 \\ \cline{3-7} 
                         &                          & \multicolumn{1}{c|}{kids} & \multicolumn{1}{c|}{bags} & \multicolumn{1}{c|}{store} & \multicolumn{1}{c|}{station} & please \\ \hline \hline
\multirow{2}{*}{Arabic}  & Train                 & \multicolumn{1}{c|}{78}   & \multicolumn{1}{c|}{78}   & \multicolumn{1}{c|}{78}    & \multicolumn{1}{c|}{78}      & 78     \\ \cline{2-7} 
                         & Test                  & \multicolumn{1}{c|}{28}   & \multicolumn{1}{c|}{28}   & \multicolumn{1}{c|}{28}    & \multicolumn{1}{c|}{28}      & 28     \\ \hline \hline
\multirow{2}{*}{French}  & Train                 & \multicolumn{1}{c|}{45}   & \multicolumn{1}{c|}{45}   & \multicolumn{1}{c|}{45}    & \multicolumn{1}{c|}{45}      & 45     \\ \cline{2-7} 
                         & Test                  & \multicolumn{1}{c|}{15}   & \multicolumn{1}{c|}{15}   & \multicolumn{1}{c|}{15}    & \multicolumn{1}{c|}{15}      & 15     \\ \hline \hline
\multirow{2}{*}{Spanish} & Train                 & \multicolumn{1}{c|}{65}   & \multicolumn{1}{c|}{65}   & \multicolumn{1}{c|}{65}    & \multicolumn{1}{c|}{65}      & 65     \\ \cline{2-7} 
                         & Test                  & \multicolumn{1}{c|}{15}   & \multicolumn{1}{c|}{15}   & \multicolumn{1}{c|}{15}    & \multicolumn{1}{c|}{15}      & 15     \\ \hline
\end{tabular}%
}
\vspace{-0.0cm}
\end{table}

For any given speech utterance, the MFCCs and FCs are extracted for the speech frames, using a framesize of 20~ms and a frameshift of 10~ms. A Mel filterbank of $K=24$ size has been used, thus resulting in a 24-dimensional feature-vector for either of the two features. Depending on the time-duration of the speech utterances, the feature-matrices will vary for either of the two features, and hence they have been fitted to a 256 $\times$ 256 matrix, with zero-padding as necessary. As such, when only MFCCs or FCs are used as features for the two-stage CNN, the input feature-set is a 256 $\times$ 256 matrix (like a greyscale image), whereas when both the features are used in combination, the input feature-set is a 256 $\times$ 256 $\times$ 2 matrix (like a colour image of two-channels only).

As mentioned earlier, for each of the three accents, a separate two-stage CNN is used to model the features for classifying the five words. The fraction of instances when the correct word has been classified provides the word recognition accuracy. The performance of the system is evaluated not only for clean speech but also for noisy speech. For this purpose, the speech utterances used for testing are corrupted with three different types of noise at different signal-to-noise-ratios (SNRs) using the NOISEX-92 corpus \cite{varga1993assessment}. Each of these types of noise test the system under different conditions. In the case of White noise, the entire spectrum of the speech signal is corrupted, whereas in the case of Babble and HFchannel noise the lower-end and higher-end of the spectrum, respectively, is mainly corrupted.

\subsection{Performance for clean speech}
\begin{table}[h!]
\vspace{-0.0cm}
\renewcommand{\arraystretch}{1.1}
\centering
\caption{ Word recognition accuracy for clean speech.}
\label{t_clean}
\resizebox{0.4\textwidth}{!}{%
\begin{tabular}{|c||ccc|}
\hline
\multirow{2}{*}{Accent} & \multicolumn{3}{c|}{Features}                                         \\ \cline{2-4} 
                        & \multicolumn{1}{c|}{MFCCs} & \multicolumn{1}{c|}{FCs}  & MFCCs ~+~FCs \\ \hline \hline
Arabic                  & \multicolumn{1}{c|}{0.78}  & \multicolumn{1}{c|}{0.48} & 0.80         \\ \hline
French                  & \multicolumn{1}{c|}{0.80}  & \multicolumn{1}{c|}{0.49} & 0.82         \\ \hline
Spanish                 & \multicolumn{1}{c|}{0.75}  & \multicolumn{1}{c|}{0.52} & 0.80         \\ \hline
\end{tabular}%
}
\vspace{-0.0cm}
\end{table}

\begin{figure*}[th!]
\centering
\includegraphics[height=9.5cm]{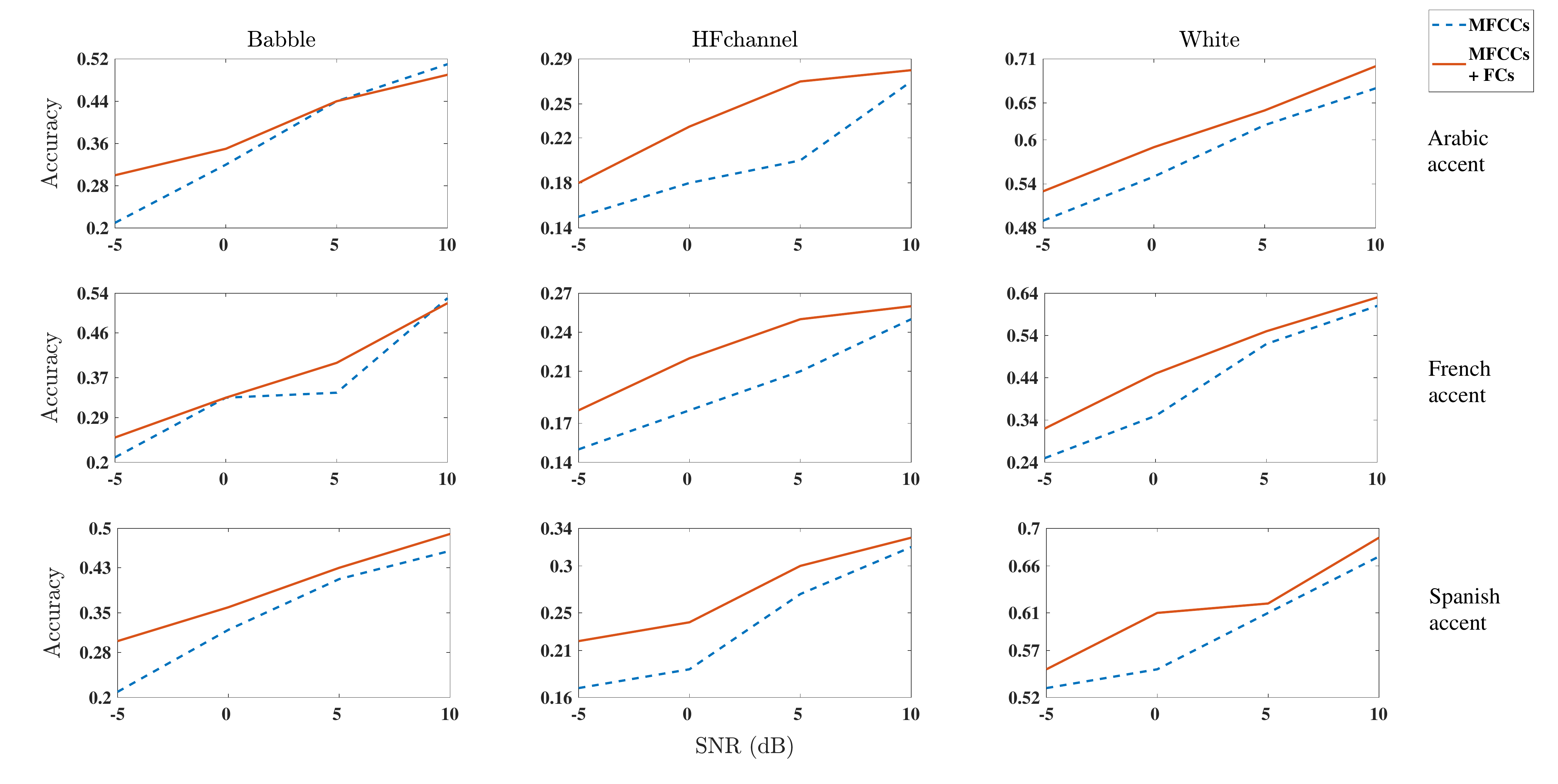}
\caption{Performance of the word recognition system for noisy speech.}
\label{f_noisy}
\end{figure*}

In Table \ref{t_clean}, the performance of the individual and combined features are presented for clean speech. As is evident, the FCs are not competitive as standalone features; however, in combination with the MFCCs, they do provide decent performance gain, for each of the three accents. 

\subsection{Performance for noisy speech}
Following the results observed in the clean speech scenario, we abstain from using the FCs as standalone features in the rest of the experiments. As such, for the noisy speech scenario, we evaluate and compare the performance of the MFCCs and the MFCCs+FCs only. Figure~\ref{f_noisy} presents the performance of the word recognition system under Babble, HFchannel, and White noise, for each of the three accents. Each row in the figure presents the performance for a particular accent. Each plot in a given row represents the performance under a particular type of noise. In each plot, there are two curves, representing the performance at different SNRs: one for the MFCCs and the other for the MFCCs+FCs combination. As is evident from the figure, irrespective of the accent, and the type and strength of noise, the proposed features (FCs), in combination with the MFCCs, provide notable performance enhancement for the AWR system. This suggests that the FCs are versatile features and consist of speech-specific information that complement the MFCCs.  

However, at this juncture, we must note that (in the case of both clean and noisy speech) we have utilized three different CNNs for the three accents. In other words, the inter-accent variation has been eliminated in our study. Had we created a single CNN model for the five words, irrespective of the accents, we may have obtained poorer performance. As such, this work is only a part of a larger project in which the accents could be identified before word recognition or an end-to-end model is developed which caters to the inter-accent variations.

\section{Conclusion and future work}
\label{sec:conclusion}

In this work, we have proposed a new set of features derived from the Mel filterbank and the Fourier spectrum of the speech signal, namely the FCs, which have been used for classifying five different English words uttered by non-native speakers of three different accents. A separate AWR system has been built for each of the three accents, but using the same features and neural-network architecture. In particular, a two-stage CNN has been proposed for this word recognition task. The results observed for both clean and noisy speech show that the FCs are versatile features which aid the MFCCs in accurate classification of the words.

The effectiveness of the FCs in the experiments conducted encourages the exploration of other kinds of spectral representation of the speech signal, such as the Hilbert Spectrum and the HNGD spectrum for extracting complementary features \cite{Huang1998,Huang2005,EMD_RS_Review,bayya2013spectro}. Further, in the limited scope of this work, the ability of the proposed features in incorporating inter-accent variation has not been studied. While the proposed features showcase their utility in the word classification task, in order to model the accent variation a larger neural network or other modelling framework needs to be employed. Lastly, since the five words used in this work were arbitrarily chosen without any acoustic or phonetic bias, it is possible that for other words the accuracy might be substantially better or worse. As such, using a larger set of words would give a better sense of the overall results. These issues may be looked into in the future.

\balance


\end{document}